\documentclass[UKenglish]{book}
\usepackage{graphicx}
\usepackage[sectionbib,sort&compress]{natbib}
\usepackage{chapterbib}
\usepackage{wileyvch}
\usepackage[breaklinks]{hyperref}
\usepackage[normalem]{ulem}

\newcommand{\om}{\omega}
\newcommand{\ra}{\rightarrow}
\newcommand{\be}{\begin{equation}}
\newcommand{\ee}{\end{equation}}
\newcommand{\reals}{\mathbb{R}}
\newcommand{\mE}{\mathcal{E}}

\DeclareMathSymbol{\jj}{\mathalpha}{letters}{"11}

\newcommand{\Jhat}{\mathbf {\hat{J}}}

\begin{document}

\author{Hugo Touchette and Rosemary J. Harris}
\title{Large deviation approach to nonequilibrium systems}

\chapterauthor{Hugo Touchette and Rosemary J. Harris}
\chapter{Large deviation approach to nonequilibrium systems}

{\setlength{\baselineskip}{1\baselineskip}
\small
\textbf{\textsf{Abstract.}} The theory of large deviations has been applied successfully in the last 30 years or so to study the properties of equilibrium systems and to put the foundations of equilibrium statistical mechanics on a clearer and more rigorous footing. A similar approach has been followed more recently for nonequilibrium systems, especially in the context of interacting particle systems. We review here the basis of this approach, emphasizing the similarities and differences that exist between the application of large deviation theory for studying equilibrium systems on the one hand and nonequilibrium systems on the other. Of particular importance are the notions of macroscopic, hydrodynamic, and long-time limits, which are analogues of the equilibrium thermodynamic limit, and the notion of statistical ensembles which can be generalized to nonequilibrium systems. For the purpose of illustrating our discussion, we focus on applications to Markov processes, in particular to simple random walks.
\par}

\section{Introduction}
\label{secintro}

Nonequilibrium systems are being increasingly studied using methods borrowed from the mathematical theory of large deviations, as developed in the 60s and 70s by Donsker and Varadhan, and Freidlin and Wentzell (see \cite{deuschel1989,dembo1998,touchette2009} for historical references). Indeed the central concepts and quantities of this theory -- e.g., the large deviation principle, rate functions, generating functions, etc. -- have now entered the standard jargon of driven nonequilibrium systems modelled as discrete- or continuous-time Markov processes (see, e.g., \cite{touchette2009,derrida2007,bertini2007}). 

With hindsight, one can argue that this evolution, although relatively recent, was to be expected: large deviation theory has been used successfully in equilibrium statistical mechanics for well over 30 years \cite{ruelle1969,lanford1973,ellis1985,touchette2009}, and so it is not surprising that this success finds its way into nonequilibrium statistical mechanics. However, there is more, in that the two scenarios -- equilibrium and nonequilibrium -- share many ideas, concepts and even a theoretical structure which happen to find a clear and precise expression in the language of large deviations. It is natural, therefore, to see this language being used for both theories.

By viewing equilibrium statistical mechanics from the point of view of large deviation theory, one gets a clear sense, for example, of why there is a Legendre transform in thermodynamics connecting the entropy and the free energy, when this Legendre transform is valid, how equilibrium states relate to the notions of concentration and typicality, and how these states arise out of variational principles such as the maximum entropy principle or the minimum free energy principle. Similar ideas and results of nonequilibrium statistical mechanics are also made clear by viewing them through the prism of large deviation theory.  In addition, one sees, as mentioned, essential similarities between equilibrium and nonequilibrium.

Our goal in this chapter is to explain these points and to illustrate them with simple examples, mainly involving continuous-time Markov processes. We start in the next section by recalling the basis and essential concepts of equilibrium statistical mechanics, and by discussing their analogues in nonequilibrium statistical mechanics. Among these, we mention the notions of statistical ensemble, stationarity, typicality, fluctuations, and scaling limit (e.g., thermodynamic limit, hydrodynamic or macroscopic limit, long-time limit). In Sec.~\ref{secldt}, we re-express these concepts in the language of large deviations to define them in a precise, mathematical way and to emphasize the theoretical structure that underlies both equilibrium and nonequilibrium statistical mechanics. We illustrate this structure with a variety of applications in Sec.~\ref{secapps} and then end in Sec.~\ref{seccon} with some concluding remarks and open problems.

Much of the large deviation content explored in the present contribution can be found with more details in \cite{touchette2009}. Here we focus on discussing the common goals, concepts, and results of equilibrium and nonequilibrium statistical mechanics, rather than providing a complete review of either subject, and on proposing a clear approach to studying nonequilibrium systems which parallels that used for studying equilibrium systems. We draw inspiration in doing so from the works of Oono \cite{oono1989,paniconi1997,oono1998}, Eyink \cite{eyink1990,eyink1996a,eyink1998}, and Maes~\textit{et al.}~\cite{maes2006,maes2008a,maes2007,maes2003a}, among others, which show the emergence of similar ideas and views as early as the late 80s.

\section{From equilibrium to nonequilibrium systems}
\label{seceqnoneq}

Before we discuss how large deviation concepts enter in equilibrium and nonequilibrium statistical mechanics, we recall in this section the basis of each theory and emphasize some concepts shared by both. Of these, the most important to keep in mind is the concept of \emph{typicality}, connected mathematically to the Law of Large Numbers and the concentration of probability distributions.

\subsection{Equilibrium systems}

The goal of \emph{equilibrium} statistical mechanics, as is well known, is to explain and predict the emergence of macroscopic equilibrium states of systems composed of many particles by treating their microscopic states in a probabilistic way. The main properties of equilibrium states are that they are stationary in time, they are stable against small perturbations, and are described by only a few variables, i.e., they are low-dimensional \emph{macro}states compared to the high-dimensional \emph{micro}states used for describing the many-particle system at the microscopic level.

Equilibrium states are also defined with respect to a given macrostate.  Physicists often say that a system is \emph{at equilibrium} or is \emph{in a state of equilibrium}, but what is meant, to be more precise, is that the system has, say, an equilibrium energy or an equilibrium magnetization. Thus an equilibrium state is a particular state or a value of a macrostate or a collection of macrostates. This is different from saying that a system \emph{is an equilibrium system}. We shall give a definition of the latter concept later, after describing the analogue of an equilibrium state for nonequilibrium systems.

For now, let us recall how equilibrium states are modelled in statistical mechanics. The basic ingredients are well known. Consider a system of $N$ particles, which for simplicity we take to be a classical system, and let the sequence $\om=(\om_1,\om_2,\ldots,\om_N)$ denote the microscopic configuration or \emph{microstate} of the system, where $\om_i$ is the state of the $i$th particle. To study the statistical properties of this system, we consider a \emph{prior} probability distribution $P(\om)$, interpreted as the stationary distribution of the microscopic dynamics.  The state space of one particle is denoted by $\Lambda$, so that $P(\om)$ is a probability distribution over the $N$-particle space $\Lambda_N=\Lambda^N$.

Next, we consider a \emph{macrostate} $M_N$, corresponding mathematically to a function $M_N(\om)$ of the microstates, and proceed to compute the probability distribution of this random variable obtained under $P(\om)$ using
\be
P(M_N=m)=\int_{\Lambda_N}\delta(M_N(\om)-m)\, P(\om)\, d\om.
\ee
If $P(\om)$ is a valid model of an equilibrium system and the macrostate $M_N$ is chosen properly, then what should be observed is that $P(M_N)$ is  concentrated around certain highly probable values and that this concentration gets more pronounced as $N$ gets larger. It is these \emph{most probable} or \emph{typical} values (points or states) of $M_N$ that we call \emph{equilibrium states} of $M_N$.

Mathematically, the concentration of $P(M_N)$ is akin to a Law of Large Numbers, in the sense that there exist sets $B$ of values of $M_N$ such that
\be
\lim_{N\ra\infty} P(M_N\in B)=1\qquad\text{and}\qquad \lim_{N\ra\infty} P(M_N\notin B)=0.
\label{eqlln1}
\ee
The smallest set $B$ having this property corresponds to the set of equilibrium values of $M_N$ or, more loosely, the set of equilibrium states of the system (as defined with respect to $M_N$). 

We shall see in the next section that an essential property of the concentration of $P(M_N)$ on $B$ is that it is \emph{exponential} as a function of $N$ (or, more generally, the volume of the system), which means that the probability that $M_N$ deviates from one of its equilibrium values is exponentially small with the system size, $N$. Physically, these deviations are termed \emph{fluctuations}, and so we say that the probability of fluctuations from equilibrium states is exponentially small with $N$.

The exponential concentration of $P(M_N)$ explains why large deviation theory is used in equilibrium statistical mechanics. Physically, it is also the reason why equilibrium states correspond to \emph{typical} values of $M_N$ and not, as often claimed, to \emph{average} values of $M_N$. The fundamental property of equilibrium states is indeed that they do not fluctuate, or at least appear not to at the macroscopic level, so the fact that $M_N$ has a well-defined average value is obviously not enough: $M_N$ must converge in probability to some typical values.

The reason why average values are used in statistical mechanics is arguably that they are conceptually simpler than typical (or concentration) values and that, if $M_N$ has a unique typical value, then its average is the same as its typical value (we say in this case that $M_N$ is \emph{self-averaging}). However, the use of averages is somewhat misleading as it detracts us from the essential property of equilibrium states, which is again that these states arise probabilistically from the concentration of a probability distribution. Equilibrium states are first and foremost \emph{typical} states arising from the scaling limit that is the \emph{thermodynamic limit} \cite{goldstein2001,goldstein2004}.

This is an important point which leads us to the discussion of extensivity versus intensivity. Physically, it should be clear that the total energy $H_N$ of an $N$-body system with short-range interactions does not concentrate in the thermodynamic limit, simply because the energy is \emph{extensive} for such a system, and so diverges with $N$. As a result, one cannot formally say that the system has an \emph{equilibrium energy} in the thermodynamic limit. Rather, the correct macrostate having an equilibrium value, i.e., the one that concentrates in the thermodynamic limit, is the energy per particle or mean energy $h_N=H_N/N$ which is an \emph{intensive} quantity.

To explain this point more clearly, let us consider a sum 
\be
S_N=\sum_{i=1}^N X_i
\ee
of $N$ independent and identically distributed random variables $X_1,\ldots,X_N$. If the mean $\langle X_1\rangle=\mu$ of these random variables is finite, then as $N\ra\infty$ the distribution of $S_N/N$ concentrates to the mean in such a way that
\be
\lim_{N\ra\infty} P(|S_N/N -\mu|>\epsilon)=0
\ee
for all $\epsilon >0$, in accordance with the Law of Large Numbers. The point to note about this result, which is equivalent to the second limit shown in (\ref{eqlln1}), is that it holds for the mean sum $S_N/N$ and not for the sum $S_N$: the distribution of $P(S_N)$ does not concentrate; in fact, it flattens as $N\ra\infty$. Similarly, it is easy to show that the distribution of $S_N/N^\alpha$ flattens for all $\alpha\in (0,1)$ and concentrates in a trivial way at zero for all $\alpha>1$. Hence, the only normalization of a  sum $S_N$ of independent (or near independent) random variables with finite mean that yields a non-trivial concentration point is $S_N/N$.

The same observation applies to equilibrium states. The reason again why equilibrium states appear stable at the macroscopic level is that the fluctuations around these states are very improbable and become the more unlikely the bigger the system gets. Mathematically, the way to make sense of this observation is to consider random variables that have the property of concentrating in the thermodynamic limit $N\ra\infty$. From this point of view, the total energy $H_N$ is not a ``good'' macrostate to consider because it does not concentrate in this limit. The same goes, similarly to $S_N$, for the macrostates $H_N/\sqrt{N}$ or $H_N/N^2$: the distribution of the former flattens, whereas the distribution of the latter concentrates trivially to zero. We get a non-trivial concentration only for $H_N/N$.

This at least is generally true for short-range interacting systems. For long-range interacting systems, such as gravitational systems or mean-field systems, the ``good'' energy macrostate to consider might be $H_N/N^2$ or, more generally, $H_N/N^\alpha$ with $\alpha\geq 2$ \cite{campa2009}. The choice of $\alpha$ will depend on the system considered, but the requirement again is that the distribution of the macrostate that is studied should concentrate when $N\ra\infty$. From this point of view, one might have to consider different thermodynamic (or scaling) limits and macrostates in order to correctly describe the equilibrium states of those systems. 

\subsection{Nonequilibrium systems}
\label{ssecnoneq}

The study of nonequilibrium systems is conceptually more difficult than that of equilibrium systems because one is interested in describing not only the stationary fluctuations, as is done for equilibrium systems, but also the fluctuating dynamics of the system arising in time and the fluctuations of macrostates or observables integrated over time. Thus, in addition to considering the number of particles present in a system (or its volume), one also needs to consider the evolution of that system in time. This implies that different scaling limits may be taken depending on the system and macrostate or observable studied. 

For definiteness, we consider here nonequilibrium systems modelled by Markovian processes. To simplify the presentation of these models, we assume for now that the stochastic evolution takes place in discrete time (although continuous-time models will also be discussed in the following sections). In this case, the \emph{microstate} $\om$ that represented before the configuration of an equilibrium system at a fixed (yet unspecified) instant of time is now a complete trajectory $\om=\{\om_i\}_{i=1}^n$ consisting of $n$ timesteps. The assumption that the process is Markovian then amounts to assuming that the prior distribution $P(\om)$ can be decomposed according to a Markov chain
\be
P(\om)=P(\om_1)P(\om_2|\om_1)\ldots P(\om_n|\om_{n-1}),
\ee
with initial distribution $P(\om_1)$ and transition matrix elements $P(\om_i|\om_{i-1})$, which, in most cases, are assumed to be time-homogeneous (i.e., time-independent). This form of prior is, from a pragmatic point of view, our stochastic model for $\om$ from which all distributions of macrostates or observables are computed. Thus, so far, the formalism is abstractly the same as for equilibrium systems: a system is described by its microstate $\om$ and a probability distribution $P(\om)$ on the space of microstates. What changes for nonequilibrium systems is the interpretation of $\om$ as a time-trajectory of a system of one or more particles.

This difference allows us to consider many types of macrostates or observables. For example, one can consider a \emph{fixed-time} or \emph{static} observable $M(\om_i)$ which is a function of the state of the system at a specific timestep $i$. One can also consider \emph{dynamic} observables of the form
\be
M_n(\om)=\frac{1}{n}\sum_{i=1}^n f(\om_i),
\label{eqaddo1}
\ee
referred to mathematically as \emph{additive observables} or \emph{additive functionals} \cite{touchette2009}, which involve states at different times. Another type of \emph{dynamic} observable, which arises in the context of particle currents, is
\be
M_n(\om)=\frac{1}{n}\sum_{i=1}^{n-1} f(\om_i,\om_{i+1}).
\label{eqcuro1}
\ee

Whatever the observable chosen, the goal when studying nonequilibrium systems is to compute the probability distribution $P(M=m)$ of a given observable $M$ starting from the prior distribution $P(\om)$ defining our model of that system, and to see whether this distribution concentrates, in some scaling limit, over specific values of $M$. The scaling limit that needs to be considered depends on the system and observable chosen: it can be the infinite-volume limit $N\ra\infty$ for a fixed-time observable $M(\om_i)$, the infinite-time limit $n\ra\infty$ for additive or current-like observables, or a combination of these two limits if the latter observables involve many particles. Other limits can also be conceived, e.g., the small-noise limit of dynamical systems perturbed by noise, the continuous-time limit of discrete-time systems, or the continuous-space limit of discrete-space systems \cite{touchette2009}. 

Some of these limits will be explored in the following sections. The essential point to note is that these \emph{scaling} or \emph{hydrodynamic} \emph{limits} are expected to give rise to a concentration of the probability distribution $P(M)$ similar to the one discussed for equilibrium systems, and thus to the emergence of typical states for the system or observable studied. 

\subsection{Equilibrium versus nonequilibrium systems}
\label{sseceqvneq}

So far we have not attempted to distinguish equilibrium from nonequilibrium systems in any precise way other than to hint that the distribution $P(\om)$ describes a ``static'' random variable in the case of equilibrium systems and a ``dynamic'' random variable, i.e., a \emph{stochastic process}, in the case of nonequilibrium systems. But what makes a system an equilibrium or a nonequilibrium system? 

To answer this question in a mathematical way, we need to consider the stochastic time evolution of a system and study how the prior distribution $P(\om)$ of its complete trajectory $\om$ behaves when the time ordering of $\om$ is reversed. To be more precise, consider the discrete-time trajectory $\om=(\om_1,\om_2,\ldots,\om_{n-1},\om_n)$ containing $n$ timesteps, and define the \emph{time-reversed} trajectory $\om^R$ associated with $\om$ as the trajectory obtained by re-ordering the states of $\om$ in reverse order, i.e., $\om^R=(\om_n,\om_{n-1},\ldots,\om_2,\om_1$). Then we say that the system modelled by $P(\om)$ is an \emph{equilibrium system} if $P(\om)=P(\om^R)$ for all $\om$ \cite{derrida2007}.\footnote{For simplicity, we assume that the $\om$'s themselves have even parity under time-reversal.} If this condition is not satisfied, then we say that the system is a \emph{nonequilibrium system}. Mathematically, this condition is equivalent to the notion of \emph{reversibility} or \emph{detailed balance}, stated here at the level of complete trajectories rather than the more usual level of transition rates. Thus, the stochastic dynamics of an equilibrium system satisfy detailed balance, whereas those of a nonequilibrium system do not. 

The rationale behind this definition is that equilibrium prior distributions, such as the microcanonical and canonical distributions, are stationary distributions of stochastic dynamics verifying detailed balance, and that the very notion of detailed balance captures the physical observation that equilibrium systems are systems in which fluctuations arise with no ``preferred'' direction in time. Nonequilibrium systems, by contrast, have a stochastic dynamics that is not symmetric under time reversal. This does not mean that nonequilibrium systems do not have stationary distributions; they often do, but the form of these distributions is generally much more complicated than equilibrium distributions.

\section{Elements of large deviation theory}
\label{secldt}

We show in this section how the mathematical theory of large deviations makes precise the observation that probability distributions of macrostates concentrate \emph{exponentially} with some scaling parameter (e.g., number of particles, volume, integration time, noise power, etc.). This exponential concentration is the source, for equilibrium systems, of the Legendre transform connecting the entropy and the free energy, and, therefore, of the Legendre structure of thermodynamics. For nonequilibrium systems, it also gives rise to a Legendre transform between quantities that are the nonequilibrium analogues of the entropy and the free energy. 

\subsection{General results}
\label{ssecgen}

To explain the central ideas and results of large deviation theory, we first consider a general random variable or macrostate $A_n$ indexed by the parameter $n$ which can, for example, be the number of particles or the number of timesteps.

The starting point of large deviation theory is the observation that the probability distribution $P(A_n)$ of $A_n$ is, for many random variables of interest, decaying to zero exponentially fast with $n$. The exponential decay is in general not exact; rather, what often happens is that the dominant term in the expression of $P(A_n)$ is a decaying exponential with $n$, so that we can write
\be
P(A_n=a)\approx e^{-nI(a)},
\ee
with $I(a)$ the rate of decay. When $P(A_n)$ has this form, we say that $P(A_n)$ or $A_n$ satisfies a \emph{large deviation principle} (LDP). To be more precise, we say that $P(A_n)$ or $A_n$ satisfies an LDP if the limit
\be
\lim_{n\ra\infty} -\frac{1}{n}\ln P(A_n=a)=I(a)
\ee
exists. The decay function $I(a)$ defined by this limit is called the \emph{rate function}. The factor $n$ in the exponential is called the \emph{speed} of the LDP \cite{ellis1985,dembo1998,touchette2009}.

The interest in large deviations arises because many random variables and stochastic processes satisfy such a principle (although not all do; see, e.g., \cite{touchette2009}). The goal of large deviation theory, in this context, is to provide methods for proving that a given random variable or process satisfies an LDP and for obtaining the rate function controlling the rate of decay of the LDP.

Among these methods, let us mention two that are especially useful. The first is known as the \emph{G\"artner-Ellis Theorem} \cite{ellis1985} and proceeds by calculating the following function:
\be
\lambda(k)=\lim_{n\ra\infty}\frac{1}{n}\ln \langle e^{nkA_n}\rangle,
\ee
known as the \emph{scaled cumulant generating function} (SCGF). The statement of the G\"artner-Ellis Theorem, in its simplified form, is that, if $\lambda(k)$ is differentiable for all $k\in\reals$, then $A_n$ satisfies an LDP with rate function given by the \emph{Legendre-Fenchel transform} of $\lambda(k)$:
\be
I(a)=\max_{k\in\reals} \{ka-\lambda(k)\}.
\ee
In many physical applications, $\lambda(k)$ is actually differentiable \emph{and} strictly convex, and in this case the Legendre-Fenchel transform reduces to the better-known \emph{Legendre transform}, written as
\be
I(a)=k(a) a-\lambda(k(a)),
\ee
with $k(a)$ the unique solution of $\lambda'(k)=a$.

The G\"artner-Ellis Theorem is useful in practice because it bypasses the direct calculation of $P(A_n)$. By calculating the SCGF of $A_n$ and by checking that this function is differentiable, we instantly prove that $P(A_n)$ satisfies an LDP and obtain the rate function controlling the concentration of $P(A_n)$ in the limit $n\ra\infty$. 

For certain random variables, $\lambda(k)$ can be calculated but is not differentiable; see \cite{touchette2009} for examples. In this case, it can be proved that the Legendre-Fenchel transform of $\lambda(k)$ yields only the convex envelope of $I(a)$ \cite{touchette2009}. To obtain the full rate function, one may use other methods, such as the \emph{contraction principle} \cite{ellis1985,dembo1998,touchette2009,touchette2010b}. The basis of this method is to express $A_n$ as a function $f(B_n)$ of some random variable $B_n$ satisfying an LDP with rate function $J(b)$, i.e.,
\be
P(B_n=b)\approx e^{-nJ(b)}.
\ee
If such a random variable and function can be found, then the contraction principle states that $A_n$ also satisfies an LDP with rate function given by
\be
I(a)=\min_{b: f(b)=a} J(b).
\label{eqcp1}
\ee

The minimization appearing in this result is a natural consequence of the approximation method known as Laplace's principle \cite{ellis1985,dembo1998,touchette2009} as applied to the integral
\be
P(A_n=a)=\int_{b:f(b)=a} P(B_n=b)\, db.
\ee
Assuming that the probability distribution $P(B_n)$ decays exponentially with $n$, this integral is dominated by the largest exponential term such that $f(b)=a$, which means that we can write
\be
P(A_n=a)\approx \exp\left({-n\min_{b:f(b)=a} J(b)}\right)
\ee
with sub-exponential correction factors in $n$. Thus we see that $P(A_n)$ satisfies an LDP with rate function given by Eq.~(\ref{eqcp1}).

The name ``contraction'' arises in the context of this result from the fact that the function $f$ can in general be a many-to-one function, in which case we are ``contracting'' the fluctuations of $B_n$ down to the fluctuations of $A_n$ in such a way that the probability of the fluctuation $A_n=a$ is the probability of the most probable (yet exponentially improbable) fluctuation $B_n=b$ leading to $A_n=a$. 

This interplay between the appearance of exponentially small terms in integrals and the possibility to approximate these integrals by their largest term using Laplace's principle also explains the appearance of the ``$\max$'' in the G\"artner-Ellis Theorem and  the Legendre transform connecting rate functions and SCGFs \cite{touchette2009}. In this sense, large deviation theory can be thought of as a ``calculus'' of exponentially decaying probability distributions, connecting the properties of integrals such as $\langle e^{nk A_n}\rangle$, which are  exponential in $n$, with the exponential properties of $P(A_n)$ itself.

\subsection{Equilibrium large deviations}
\label{sseqld}

The application of the results stated above to equilibrium systems is straightforward. For definiteness, we consider a general macrostate $M_N(\om)$ involving $N$ particles and study its probability distribution $P_\beta(M_N)$ in the canonical ensemble defined by the prior probability distribution
\be
P_\beta(\om)=\frac{e^{-\beta H_N(\om)}}{Z_N(\beta)},\qquad Z_N(\beta)=\int_{\Lambda_N} e^{-\beta H_N(\om)}\, d\om,
\ee
where $H_N$ is the Hamiltonian of the system considered.

If $P_\beta(M_N)$ satisfies an LDP, then the limit
\be
\lim_{N\ra\infty } -\frac{1}{N}\ln P(M_N=m)=I_\beta(m)
\ee
exists and defines the rate function $I_\beta(m)$ of $M_N$ in the canonical ensemble at fixed inverse temperature $\beta$. The SCGF associated with this rate function is
\be
\lambda_\beta(k)=\lim_{N\ra\infty} \frac{1}{N}\ln \langle e^{Nk M_N}\rangle_\beta,
\ee
where
\be
\langle e^{NkM_N}\rangle_\beta=\int_{\Lambda_N} e^{NkM_N(\om)} P_\beta(\om)\, d\om.
\ee
If $\lambda_\beta(k)$ is differentiable in $k$, then by the G\"artner-Ellis Theorem we have that $I_\beta(m)$ is the Legendre-Fenchel transform of $\lambda_\beta(k)$.

The connection between the LDP of $M_N$ and its equilibrium values comes directly by writing the LDP informally in the form
\be
P_\beta(M_N=m)\approx e^{-N I_\beta(m)}.
\ee
Since rate functions are always positive, this result shows that $P_\beta(M_N)$ decays exponentially fast with $N$ except at points where $I_\beta(m)$ vanishes. As noted before, these points must correspond to the points where $P_\beta(M_N)$ concentrates in the limit $N\ra\infty$, and so to the equilibrium values of $M_N$. Mathematically, we therefore define the set $\mE_\beta$ of equilibrium values of $M_N$ in the canonical ensemble as the set of global minima and zeros of the rate function $I_\beta(m)$:
\be
\mE_\beta=\{m:I_\beta(m)=0\}.
\ee
A similar definition can be given for the equilibrium values of $M_N$ in the microcanonical ensemble or any other ensemble by replacing $P_\beta(\om)$ with the prior probability distribution defining these ensembles; see Sec.~5.3 of \cite{touchette2009}. 

The rate function describes of course not only the equilibrium states but also the fluctuations around these states. In particular, if the rate function $I_\beta(m)$ admits a Taylor expansion of the form
\be
I_\beta(m)=a (m-m^*)^2 +O(|m-m^*|^3)
\ee
around a given equilibrium value $m^*$, then the \emph{small} fluctuations of $M_N$ around $m^*$ are Gaussian-distributed. The rate function, however, is rarely an exact parabola which means that the \emph{larger} fluctuations of $M_N$ away from $m^*$ are in general not Gaussian-distributed. Their distribution is determined by the rate function $I_\beta(m)$ and depends on the system studied.

This explains the word ``large'' in large deviation: contrary to the Central Limit Theorem, which gives only information about the distribution of random variables around their mean, large deviation theory gives information about this distribution near the mean but also away from the mean -- i.e., it gives information about both the \emph{small} and the \emph{large} fluctuations or deviations of random variables. From this point of view, large deviation theory can be thought of as generalizing both the Law of Large Numbers and the Central Limit Theorem.

These observations are valid for any random variable. A more specific connection with equilibrium systems can be established by studying the large deviations of the mean energy $h_N=H_N/N$ with respect to the uniform distribution $P(\om)=1/|\Lambda_N|$. In this case, the integral 
\be
P(h_N=u)=\int_{\Lambda_N}\delta(h_N(\om)-u)\, P(\om)\, d\om
\ee
is, up to a multiplicative constant, the density of states giving the number of microstates having a given mean energy $h_N(\om)=u$. For most (if not all) equilibrium systems, the density of states is known to grow exponentially with $N$ (or, more generally, the volume) and this directly implies an LDP for $P(h_N=u)$, which we write as
\be
P(h_N=u)\approx e^{N s(u)}.
\ee 
In this form, it is clear that the function $s(u)$ obtained with the limit
\be
s(u)=\lim_{N\ra\infty}\frac{1}{N}\ln P(h_N=u),
\ee
is the thermodynamic entropy associated with $h_N$. To be more precise, it is the entropy of $h_N$, as usually calculated from the density of states, minus an unimportant additive constant; see Sec.~5.2 of \cite{touchette2009}.

To complete the connection with thermodynamics, note that the generating function 
\be
\langle e^{N k h_N}\rangle=\int_{\Lambda_N} e^{N k h_N(\om)}\, P(\om)\, d\om
\ee
can be interpreted as the canonical partition function, whereas the SCGF of $h_N$, defined as
\be
\lambda(k)=\lim_{N\ra\infty}\frac{1}{N}\ln \langle e^{N k h_N}\rangle,
\ee
can be seen as the analogue of the free energy function of the canonical ensemble. To be more precise, re-define the partition function $Z_N(\beta)$ by including the prior uniform distribution $P(\om)$ in the integral over $\Lambda_N$:
\be
Z_N(\beta)=\int_{\Lambda_N} e^{-\beta H_N(\om)}\,P(\om)\, d\om
\ee
and define the free energy\footnote{In thermodynamics, the free energy is more commonly defined with an additional factor $1/\beta$ in front of the logarithm.} by
\be
\varphi(\beta)=\lim_{N\ra\infty}-\frac{1}{N}\ln Z_N(\beta).
\ee
Then, it is easy to see that $\varphi(\beta)=-\lambda(k)$ with $k=-\beta$. From this connection, it is also easy to see that the G\"artner-Ellis Theorem implies that, if $\varphi(\beta)$ is differentiable, then 
\be
s(u)=\min_{\beta\in\reals} \{\beta u-\varphi(\beta)\}.
\ee
This Legendre-Fenchel transform, with its inverse transform expressing $\varphi(\beta)$ in terms of $s(u)$ (see Sec.~5.2 of \cite{touchette2009}), is the formal expression of the Legendre transform appearing in thermodynamics. The large deviation derivation of this transform makes it clear that it is valid under specific mathematical conditions (viz., the differentiability of $\varphi$) and that it arises because of the exponential nature of both $P(h_N)$ and $Z_N(\beta)$ and the Laplace's principle linking these two functions in the thermodynamic limit. In this sense, the Legendre transform of thermodynamics does not arise out of any physical requirement -- it is a consequence of the large deviation structure of statistical mechanics, which appears in the thermodynamic limit.

\subsection{Nonequilibrium large deviations}
\label{ssecneqld}

Let us now consider a nonequilibrium macrostate or observable $M_n(\om)$ involving $n$ timesteps of a Markov process $\om=(\om_1,\om_2,\ldots,\om_n)$ described by the transition matrix elements $P(\om_{i}|\om_{i-1})$. The large deviation properties of $M_n$ can be studied similarly as done for equilibrium macrostates by calculating the SCGF $\lambda(k)$ associated with $M_n$ and by obtaining the rate function of $M_n$ as the Legendre-Fenchel transform of $\lambda(k)$ provided that $\lambda(k)$ is differentiable.

The Markov structure of the process underlying $M_n$ can be used here to obtain more explicit expressions for $\lambda(k)$. In the case of the additive observable shown in Eq.~(\ref{eqaddo1}), for example, we have
\be
\lambda(k)=\ln \zeta(\mathbf{P}_k),
\ee
$\zeta(\mathbf{P}_k)$ being the largest eigenvalue of the so-called \emph{tilted} transition matrix $\mathbf{P}_k$ with elements
\be
P_k(\om_{i}|\om_{i-1})=P(\om_i|\om_{i-1})e^{k f(\om_i)}.
\ee
For the current-like observable $M_n$ shown in Eq.~(\ref{eqcuro1}), we have the same result but with $\mathbf{P}_k$ now given by
\be
P_k(\om_i|\om_{i-1})=P(\om_i|\om_{i-1}) e^{k f(\om_i,\om_{i-1})}.
\ee
These results are valid if the state-space of the Markov chain is bounded. For unbounded state-spaces, $\lambda(k)$ is not necessarily given by the logarithm of the dominant eigenvalue of $\mathbf{P}_k$. We shall see a related example in Sec.~\ref{secsubt}.

For Markov processes evolving continuously in time, the above statements translate into the following ones. For an additive functional of the form
\be
M_T(\om)=\frac{1}{T}\int_0^T f(\om_t)\, dt,
\ee
the SCGF $\lambda(k)$, calculated in the limit $T\ra\infty$, is given by the largest eigenvalue of the \emph{tilted generator},
\be
G_k(\om',\om)=G(\om',\om)+kf(\om) \delta_{\om',\om},
\ee
with $G(\om',\om)$ the elements of the generator of the original process.  Note the absence of the logarithm here, as we are dealing with the generator, not the transition matrix.  For current-like observables having the form  
\be
M_T(\om)=\frac{1}{T}\sum_{i=0}^{N(T)-1} f(\om_{t_i},\om_{t_{i+1}}) \label{eqcuro2}
\ee
where the sum is over the random transitions between states happening at times $\{t_0,t_1,\ldots,t_{N(T)-1}\}$, $\lambda(k)$ is given instead by the largest eigenvalue of a tilted generator with elements
\be
G_k(\om',\om)=G(\om',\om)e^{kf(\om,\om')}. 
\label{e:currgen}
\ee

Observables involving other scaling limits, in addition to the time or particle number limits, can be treated in a similar way, as discussed for example in Sec.~\ref{ss:macro}. In all cases, the LDPs that we obtain give us information about the fluctuations of the observable of interest similar to the information obtained from equilibrium LDPs. In particular, the global minima and zeros of the rate function determine the typical values of that observable, which are physically interpreted as typical steady states or hydrodynamic states or equations, depending on the observable studied. Then, as for random variables in general, the shape of the rate function around its minima determines the behaviour of the small and large fluctuations of the observable around its typical values.

With the knowledge of the rate function of an observable, it is possible for example to determine whether this observable satisfies the so-called fluctuation relation symmetry. Consider, to be specific, an observable $M_T$ integrated over the time $T$ and assume that $M_T$ satisfies an LDP with rate function $I(m)$.  We say that $M_T$ satisfies a \emph{Gallavotti-Cohen-type fluctuation relation} if 
\be
\frac{P(M_T=m)}{P(M_T=-m)}\approx e^{T c  m}, \label{e:MGCFT}
\ee
with $c$ a positive constant. This means that the positive fluctuations of $M_T$ are exponentially more probable than negative fluctuations of equal magnitude. In large deviation terms, it is easy to see that a sufficient condition for having this result is that $I(m)$ satisfy the following symmetry relation: 
\be
I(-m)-I(m)=cm.
\ee
In terms of the SCGF, we have equivalently
\be
\lambda(k)=\lambda(-k-c).
\ee
The next section includes an example of a very simple Markov process having this fluctuation symmetry. For other more complicated examples, see, e.g., \cite{Me07,touchette2009}.

\section{Applications to nonequilibrium systems}
\label{secapps}

In this section we aim to illustrate, more concretely, how the large deviation formalism of the preceding section can be applied to nonequilibrium systems as introduced in Sec.~\ref{ssecnoneq}. We concentrate on Markov processes in continuous time, providing a detailed pedagogical treatment for toy models of random walkers and indicating how the same techniques can be applied to more complicated (and hence more interesting) models of interacting particles.  Among other sources we draw here on the comprehensive review of Derrida~\cite{derrida2007} in which further details of many-particle applications can be found.

\subsection{Random walkers in discrete and continuous time}
\label{ss:rw}

To fix ideas, let us start by analysing perhaps the simplest possible model -- a random walker in discrete space.  Specifically, we consider a particle moving on a one-dimensional lattice of $L$ sites with, for now, periodic boundary conditions.
The microstate of the model is the particle's position $\om\in\{1,2,\ldots, L\}$ and, in discrete-time, the probabilities to move between those positions  $P(\om_i|\om_{i-1})$ are contained in the transition matrix $\mathbf{P}$.  We assume that the particle has probability $p $ per timestep, with $0<p<1$, to move in the clockwise direction and probability $q$ per timestep, with $0<q\leq 1-p$, to move in the anti-clockwise direction. Note that, if $p + q  < 1$, the particle also has a finite probability to remain stationary in a given timestep.
The position of the particle on the state space $\{1,2,\dots,L\}$ is thus a Markov chain with transition matrix 
\begin{equation}
\mathbf{P}=
\begin{pmatrix}
1-p - q  & p  & 0 & \dots & q  \\
q & 1-p - q  & p  & \dots & 0 \\
0 & q & 1-p - q  & \dots & 0\\
\vdots & \vdots & \vdots & \ddots & \vdots \\
 p  & 0 & 0 & \dots &  1-p - q  
\end{pmatrix}.
\end{equation}

It is a simple exercise to show that this Markov chain has a limiting (stationary) distribution with probability $1/L$ for the particle to be found on any given site -- an intuitively obvious result!  Slightly more interesting, from the nonequilibrium point of view, is the particle current $J_n(\om)$  which we define as the net number of clockwise jumps made by the particle \emph{per timestep}. This is a function of the form~\eqref{eqcuro1} with
\be
f(\om_i,\om_{i+1})= \delta_{\om_i+1,\om_{i+1}}-\delta_{\om_i-1,\om_{i+1}}.
\ee

A straightforward calculation shows that the mean stationary current is given by $\langle J_n \rangle = p-q$. We shall see in Sec.~\ref{ss:curr} that this corresponds to the concentration point of the probability distribution $P(J_n=j)$ which satisfies an LDP. For now, note that there is an obvious qualitative difference between the case of $p=q$ (zero mean current) and the case of $p\neq q$ (non-zero mean current).  Mathematically, this difference is just the distinction between a reversible and a non-reversible Markov chain, in the sense of detailed balance. As explained in Sec.~\ref{sseceqvneq}, we identify the former case with an equilibrium system and the latter with a nonequilibrium system.
 
The continuous-time version of this random walk can be understood by associating a physical time increment $\Delta t$ with each discrete timestep (where above we implicitly assumed $\Delta t=1$), setting the hopping probabilities per timestep to $p \Delta t$ and $q \Delta t$ and then taking the limit $\Delta t \to 0$.  Formally, our particle then remains at a given site for an exponentially distributed waiting time, with mean $1/(p+q)$, before moving clockwise, with probability $p/(p+q)$, or anti-clockwise, with probability $q/(p+q)$. Note, in particular, that $p$ and $q$ are now interpreted as rates rather than probabilities and can each be greater than unity. The infinitesimal generator corresponding to this process is 
\begin{equation}
\mathbf{G}=
\begin{pmatrix}
-p - q  & p  & 0 & \dots & q  \\
q & -p - q  & p  & \dots & 0 \\
0 & q & -p - q  & \dots & 0\\
\vdots & \vdots & \vdots & \ddots & \vdots \\
 p  & 0 & 0 & \dots & -p - q  
\end{pmatrix}.
\end{equation}

Unsurprisingly, a picture similar to the discrete-time case emerges: the process has a stationary state which has mean density $1/L$ on each site and mean current $p-q$.  Following the previous sections, we shall be interested next in deriving such mean values as concentration points of LDPs. To be specific, we shall illustrate the general discussion below with explicit calculations related to the continuous-time random walk model and various modifications of it.   We note that, as mentioned in Sec.~\ref{ssecnoneq}, the appropriate scaling limit for which an LDP holds depends on the observable we wish to consider. 

\subsection{Large deviation principle for density profiles}
\label{ss:dens}

A general interacting particle system has a state space consisting of all possible particle configurations and, on a coarse-grained scale, one is often interested in the probability of observing a particular fixed-time density profile in space.  This leads to the concept of a density function LDP which can be straightforwardly extended from equilibrium to nonequilibrium; see e.g.,~\cite{derrida2007}. 

The appropriate scaling limit expressing such an LDP is the infinite volume (and infinite particle number) limit. To be precise, for a system defined on a lattice of linear size $L$ in $d$ dimensions, one considers taking the thermodynamic limit $L \to \infty$ whilst rescaling the coordinates $\mathbf{r}$ to  $\mathbf{x}=\mathbf{r}/L$. The probability of seeing a given density profile $\rho(\mathbf{x})$ is then expected to obey
\begin{equation}
P[\rho(\mathbf{x})] \approx  \exp\left\{-L^d \mathcal{F}[\rho(\mathbf{x})]\right\}
\end{equation}
as $L\ra\infty$. This is a \emph{functional LDP}, as $P$ and $\mathcal{F}$ are both functionals of $\rho(\mathbf{x})$.\footnote{$\mathcal{F}$ is the large deviation analogue of the Ginzburg-Landau free energy expressed as a function of the particle density in the grand-canonical ensemble.} The use of the square brackets emphasizes this point.

For equilibrium systems with short-range interactions, the form of the large deviation rate functional $\mathcal{F}$ is obtained from the knowledge of $f(\rho)$, the free energy per site, as
\begin{equation}
\mathcal{F}[\rho(\mathbf{x})] = \int_0^1 \left[ f(\rho(\mathbf{x}))-f(\rho^*)-(\rho(\mathbf{x})-\rho^*)f'(\rho^*) \right] \, d\mathbf{x},
\end{equation}
where $\rho^*$ is just the mean number of particles per site~\cite{derrida2007}.  We see here that $\mathcal{F}=0$ for the uniform density profile $\rho(x)=\rho^*$.  In other words, as expected, $\rho^*$ is the typical density about which the probability distribution concentrates in the large volume limit.

To illustrate these results, let us consider a collection of independent random walkers in one dimension, discussed as a special case in \cite{Derrida09c}.  To facilitate later generalizations, we now choose to work with open boundaries rather than periodic boundary conditions. Specifically, we modify our set-up to consider $L$ sites coupled to left and right boundary reservoirs with densities $\rho_L$ and $\rho_R$ respectively, so that particles are input from the left reservoir with rate $p\rho_L$ and from the right reservoir with rate $q\rho_R$. In the bulk, and for exiting the system, each particle independently has the dynamics of the single random walker defined in Sec.~\ref{ss:rw} above. 

For equal reservoir densities, $\rho_L=\rho_R=\rho^*$, it is a relatively simple exercise to show that for any $p$, $q$, the number of particles on each site is a Poisson distribution with mean $\rho^*$.   The corresponding free energy (see e.g.,~\cite{Derrida09c}) leads to a large deviation functional of the form 
\begin{equation}
\mathcal{F}[\rho(\mathbf{x})]= \int_0^1 \left[ \rho^*-\rho(x)+\rho(x)\ln\frac{\rho(x)}{\rho^*} \right] \, dx.
\end{equation}
A Taylor expansion of the integrand readily demonstrates Gaussian fluctuations about  $\rho(x)=\rho^*$. Note that, in this special case of equal reservoir densities, the form of $\mathcal{F}$ is the same for $p=q$ and $p\neq q$: it is a local, convex, functional as is typically the case for equilibrium. Furthermore, an ensemble equivalence argument suggests that in the $L \to \infty$ limit, the density large deviation functional would be the same for periodic boundary conditions with a fixed mean density $\rho^*$. 

For interacting particle systems with \emph{non-equal} reservoir densities (i.e., boundary driving) the situation is much more interesting.  In particular, the density large deviation functional is generically expected to have a non-local structure reflecting the long-range spatial correlations characteristic of nonequilibrium. This is seen for example, in the case of the symmetric simple exclusion process~\cite{Derrida01,Derrida02b} which can be treated analytically by using the well-known matrix product ansatz~\cite{Derrida93b} and an associated additivity property.  In the corresponding asymmetric simple exclusion process, $\mathcal{F}$ is non-convex for some parameters indicating a phase transition~\cite{Derrida02,Derrida03}.  The form of $\mathcal{F}$ in certain models can be obtained by utilising the macroscopic fluctuation theory of Bertini \emph{et al.}~\cite{Bertini02,bertini2007} to which we shall return in Sec.~\ref{ss:macro}.  

\subsection{Large deviation principle for current fluctuations}
\label{ss:curr}

The presence of non-zero currents is a generic feature of nonequilibrium stationary states. For Markov processes, one generically finds that a given current $J_T$ (e.g., the net number of particles hopping between two lattice sites) time-averaged over the interval $[0,T]$ obeys a large deviation principle with speed $T$, i.e., 
\begin{equation}
P(J_T=j) \approx  e^{-TI(j)}.
\end{equation}
The relevant scaling limit here is the long-time, $T \to \infty$, limit. Although this may be combined with an infinite volume limit (as in Sec.~\ref{ss:macro} below), there is particular interest in current fluctuations in \emph{small systems}, e.g., trapped colloidal particles (see the chapter of this book by Reid~\emph{et al.}) or single molecule biological experiments (see the contribution by Alemany~\emph{et al.}). In this spirit, we use this section to illustrate the calculation and properties of $I(j)$ for a single random walker. For a treatment of general Markov diffusions, see \cite{maes2008a,maes2008}.

As follows from the general discussion in Sec.~\ref{ssecneqld}, the rate function $I(j)$ can be obtained as the Legendre-Fenchel transform of the SCGF $\lambda(k)$ of $J_T$, which, for a continuous-time process with finite state-space, is given by the principal eigenvalue $\zeta(k)$ of a tilted generator.  For example, for the single-particle random walker on a ring 
with a current $J_T$ defined, as in Sec.~\ref{ss:rw}, to be the net number of clockwise jumps the particle makes per unit time, then we need the principal eigenvalue of the matrix 
\begin{equation}
\mathbf{G}(k)=
\begin{pmatrix}
-p - q  & pe^{k}  & 0 & \dots & qe^{-k}  \\
qe^{-k} & -p - q  & pe^{k}  & \dots & 0 \\
0 & qe^{-k} & -p - q  & \dots & 0\\
\vdots & \vdots & \vdots & \ddots & \vdots \\
 pe^{k}  & 0 & 0 & \dots & -p - q  
\end{pmatrix}.
\end{equation}
It is easy to show that the normalized vector $(1/L, 1/L, \dots, 1/L)$ is a left-eigenvector of this matrix with eigenvalue
\begin{equation}
-p(1-e^{k})-q(1-e^{-k})
\end{equation} 
and an appeal to Perron-Frobenius theory~\cite{Minc88} supports the assertion that this is the desired principal eigenvalue $\zeta(k)$ which is equal to the SCGF $\lambda(k)$. From here it is a straightforward, albeit tedious, exercise to calculate $I(j)$:
\be
I(j)= \max_k \{ k j - \lambda(k) \} = p + q - \sqrt{j^2 + 4pq} + j \ln\frac{j +\sqrt{j^2 + 4pq}}{2p}.
\ee

Note that this result can also be obtained by arguing that the clockwise jumps form a Poisson process with rate $p$, whereas the anti-clockwise jumps form a Poisson process with rate $q$. Considering the long-time limit of the Poisson process, we hence have separate large deviation functions for the clockwise current $J_+$ and the anti-clockwise current $J_-$ with respective rate functions
\begin{equation}
I(j_+)= p - j_+ + j_+\ln\frac{j_+}{p}, \quad\quad I(j_-)= q - j_- + j_-\ln\frac{j_-}{q}.
\end{equation}
The rate function for the net current $J_T=J_+-J_-$ can then be obtained by the method of contraction discussed in Sec.~\ref{ssecgen}.

Notice that the rate function for the current of the random walker obeys the fluctuation relation symmetry
\be
I(-j)-I(j)=cj
\ee
with $c=\ln(p/q)$, so that
\begin{equation}
\frac{P(J_T=j)}{P(J_T=-j)}=e^{Tcj}.\label{e:FT}
\end{equation}
This result is a simple example of a fluctuation relation of the Gallavotti-Cohen type~\cite{Evans94,Gallavotti95, Lebowitz99}. As mentioned in Sec.~\ref{ssecneqld}, this result can also be expressed as the SCGF property $\lambda(k)=\lambda(-k-c)$
which derives ultimately from a straightforwardly-verified symmetry of the tilted generator:
\be
\mathbf{G}(k)^T=\mathbf{G}(-k-c) \label{e:sym}
\ee
where $T$ denotes here transpose (not time). Note that for other currents (e.g., counting the jumps across just a single bond), $\mathbf{G}(k)^T$ is no longer equal to $\mathbf{G}(-k-c)$ but, under quite general conditions, is related to it by a similarity transform so that the two matrices have identical eigenvalues and the symmetry~\eqref{e:FT} still holds. The diagonal change-of-basis matrix in the similarity transform is related to current boundary terms which, for finite state-space, are irrelevant in the long-time limit. The plethora of different finite-time fluctuation relations can be associated with different choices for these boundary terms; see, e.g.,~\cite{Me07, Rakos08} and elsewhere in this volume.

We shall see with an example below that, for infinite state space, the boundary terms may become relevant.
In passing, we also note that whilst the fluctuation theorem can be elegantly expressed as a property of the large deviation rate function, the existence of a large deviation principle is not, as sometimes believed, a \emph{necessary} prerequisite for the existence of a fluctuation relation of form~\eqref{e:FT}. A simple counter-example is provided by a random walker with right and left hopping rates increasing in time as $p\times t$ and $q \times t$ respectively.  It is easy to show that such a system has no stationary state (the mean current increases indefinitely), but since the ratio of rates of right and left steps is constant, a relation of the form~\eqref{e:FT} still holds.

\subsection{Interacting particle systems: features and subtleties}
\label{secsubt}

Thus far, the explicit examples of this section have been concerned with single-particle random walks or non-interacting collections thereof.  The same general formalism applies for interacting particle systems~\cite{Liggett85}, although analytically tractable models are the exception rather than the rule.  Paradigmatic examples include the symmetric and asymmetric simple exclusion processes, mentioned already in Sec.~\ref{ss:dens}, and the zero-range process (ZRP)~\cite{Evans05}.  Among other results, the current large deviations in the open-boundary asymmetric exclusion process have recently been calculated~\cite{Lazarescu11b,gier2011}.  Here we concentrate on the ZRP with open boundary conditions~\cite{Levine04c} (connected to so-called Jackson networks of queueing theory~\cite{Jackson57}) in order to exemplify subtleties arising in systems with unbounded state space.

For our purposes, it suffices to consider the ZRP on a one-dimensional open lattice, although related issues have also been examined for queuing models on more complicated geometries~\cite{Chernyak10d}. In one dimension, each site $l \in \{1,2,\ldots,L\}$ contains an integer number of particles $n_l$ which can hop to the nearest neighbour sites according to a continuous-time dynamics. Specifically, in the bulk the topmost particle on each site hops to the right (or to the left) with rate $p w_{n}$ (respectively, $q w_{n}$) where $w_{n}$ is a function of the number of particles $n$ on the departure site.  Particles are injected onto site $1$ (or $L$) with rate $\alpha$ (respectively, $\delta$) and extracted with rate $\gamma w_{n}$ (respectively, $\beta w_{n}$).

The properties of the model depend crucially on the function $w_n$. The choice $w_n \propto n$ corresponds to non-interacting particles, such as the random walkers considered above, whereas other forms represent an effective on-site attraction or repulsion.  In particular, if $w_n$ is bounded as $n \to \infty$, i.e.,
\be
 \lim_{n\to\infty} w_n=a<\infty,
\ee
then the model exhibits a condensation transition where, for some choices of boundary rates, particles ``pile up'' indefinitely at one or more sites and there is no stationary state.   For boundary rates outside this regime, the stationary state of the model is a product measure characterized by a site-dependent fugacity.  The mean current across each bond (i.e., between each pair of sites) depends on the rates $p$, $q$, $\alpha$, $\beta$, $\gamma$, $\delta$, but not explicitly on $w_n$. However, the form of $w_n$ determines the relationship between the fugacity and the particle density and also the relationship between $\alpha$ and $\delta$ and effective reservoir densities at the boundaries. 

To calculate the fluctuations around the stationary-state current, we can try to look, as above, for the principal eigenvalue of a tilted generator (which can be represented in terms of tensor products of matrices encoding the particle dynamics on each site). The form of this tilted generator will depend on the bond(s) across which we choose to measure the current. In the case where $w_n$ is unbounded, in the sense that $w_n \to \infty$ as $n \to \infty$, then the spectrum of the tilted generator is always gapped and the Legendre transform of the principal eigenvalue, which can be explicitly calculated in terms of the transition rates, gives the large deviation rate function for all values of current.  Furthermore, as might be expected, the principal eigenvalue, and hence the current fluctuations, are the same for currents across all bonds.

On the other hand, for $w_n$ bounded, the spectrum of the tilted generator becomes gapless for some values of $k$ and certain boundary terms can also diverge. Mathematically, this means that $\lambda(k)$ is no longer simply given by the principal eigenvalue. Physically, this possibility is related to the fact that, over long-but-finite timescales, an arbitrarily large number of particles can accumulate on each site.  This manifests in the following properties of the current large deviation function:
\begin{itemize}
\item It is bond inhomogeneous so that the probability of seeing extreme current fluctuations depends on where the current is measured;
\item It depends on the initial probability distribution of the system;
\item It does not obey the Gallavotti-Cohen fluctuation relation for large current fluctuations.
\end{itemize}

All of these properties are seen even in the single site zero-range process, for which the complete spectrum of the tilted generator and the form of $\lambda(k)$ for all $k$ can be explicitly obtained~\cite{Me06b,Rakos08}.  A phase diagram in $x$--$k$ space, where $x$ is the fugacity characterizing the initial state, reveals that there are two types of ``phase transition'' in $\lambda(k)$ analogous to first-order and continuous transitions in equilibrium. At the former, $\lambda(k)$ has a non-differentiable point, while at the latter, it remains differentiable.  An attentive reader may question how we can then obtain $I(j)$, since the G\"artner-Ellis Theorem of Sec.~\ref{ssecgen} requires differentiability of $\lambda(k)$ for all $k$. In fact, the Legendre-Fenchel transform yields the convex envelope of $I(j)$ which contains linear sections corresponding to the non-differentiable points of $\lambda(k)$. For the ZRP, one can argue on physical grounds that this \emph{is} the correct form of $I(j)$ because the most probable way to realise an average current $j$ in the linear regime involves a phase separation in time with the system spending part of its history in a state with one average current and part in a state with another average current.  This argument relies on the system having only short-range correlations in time (just as the analogous Maxwell construction in equilibrium requires short-range correlations in space) so it might be expected to fail, for example, in non-Markovian systems. 

\subsection{Macroscopic fluctuation theory}
\label{ss:macro}

Underlying the so-called \emph{macroscopic fluctuation theory} is the concept of the hydrodynamic limit which describes the emergence of a deterministic coarse-grained description from stochastic microscopic rules~\cite{Kipnis99}. Recall from Secs.~\ref{seceqnoneq} and~\ref{secldt} that such a non-fluctuating macroscopic state, corresponding to the concentration point of some probability distribution, is expected to be given by the zero of a large deviation rate function.  In this subsection, we sketch the approach of Bertini~\emph{et~al.}~\cite{Bertini02,bertini2007} for calculating this rate function.

We are interested here in systems with particle conservation in the bulk and a key ingredient is the functional form of the dependence of the instantaneous local current on the density. The correct scaling required so that the \emph{joint} distribution of current and density profiles concentrates in the limit $L\to \infty$ depends on the form of this current-density relationship.  Specifically, we focus our attention here on \emph{diffusive} processes for which the relevant macroscopic coordinates are $\mathbf{x}=\mathbf{r}/L$ and $\tau=t/L^2$. This class of systems includes symmetric and weakly-asymmetric versions of both the exclusion process and the zero-range process, but not their asymmetric counterparts for which \emph{Euler scaling} $\tau=t/L$ is needed. To illustrate loosely the procedure involved in taking the hydrodynamic limit, we now return to our favourite example of random walkers.

Consider a collection of non-interacting particles on a one-dimensional lattice with boundary reservoirs $\rho_L$ and $\rho_R$, as in Sec.~\ref{ss:dens}, and a weak asymmetry in the bulk hopping dynamics, viz., $p=1/2+E/2L$ and $q=1/2-E/2L$.  The starting point for the hydrodynamic description is to consider the mean current between two neighbouring lattice sites, $l$ and $l+1$, say.  In terms of the site densities (mean occupation numbers) $\rho_l$ and $\rho_{l+1}$, the mean current is
\be
\langle J_{l,l+1} \rangle = p \rho_{l}-q \rho_{l+1},
\ee
which yields a lattice continuity equation of the form
\be
\frac{\partial \rho_l (t)}{\partial t} =  \langle J_{l-1,l} \rangle - \langle J_{l,l+1} \rangle
= [p \rho_{l-1}(t)-q \rho_l(t)]-[p \rho_{l}(t)-q \rho_{l+1}(t)].
\ee
Now writing $t=L^2 \tau$ and $l=xL$ we have
\begin{equation}
\frac{1}{L^2} \frac{\partial \rho(x,\tau)}{\partial \tau} =  \left[p \rho(x-\tfrac{1}{L},\tau)-q \rho(x,\tau)\right]-\left[p \rho(x,\tau)-q \rho(x+\tfrac{1}{L},\tau)\right].
\end{equation}
Then assuming \emph{local stationarity} and carrying out a Taylor expansion to second order, yields
\begin{equation}
\frac{\partial \rho(x,\tau)}{\partial \tau} = -\frac{\partial}{\partial x} \left[ E \rho - \frac{1}{2} \frac{\partial \rho}{\partial x} \right] \equiv  -\frac{\partial}{\partial x} \hat{J}(x,\tau)
\end{equation}
with $\hat{J}(x,\tau)$ a rescaled current.  

Similarly, for general $d$-dimensional diffusive systems obeying Fick's Law at the macroscopic level, we expect 
\begin{equation}
\Jhat (\mathbf{x},t)= \sigma [\rho(\mathbf{x},t)] \mathbf{E} - D[\rho(\mathbf{x},t)]\mathbf{\nabla} \rho(\mathbf{x},t)  
\label{e:curr1}
\end{equation}
where $D[\rho(\mathbf{x},t)]$ is the diffusivity associated with the density profile $\rho(\mathbf{x},t)$ and $\sigma [\rho(\mathbf{x},t)]$ is the corresponding mobility.  The hydrodynamic equation, describing the deterministic or macroscopic limit as $L \to \infty$, is
\begin{equation}
\frac{\partial \rho(\mathbf{x},\tau)}{\partial \tau} = - \mathbf{\nabla} \cdot\Jhat (\mathbf{x},\tau).
\label{e:cont}  
\end{equation}

Together, Eq.~\eqref{e:curr1} and \eqref{e:cont} only represent the macroscopic or typical behavior obtained in the hydrodynamic limit. To describe the fluctuations around this limit, let us now find the joint rate function of the density and current. Specializing to the case $\textbf{E}=0$,  one observes (see, e.g., \cite{derrida2007}) that for boundary reservoirs with equal density $\rho^*$ the fluctuations of the microscopic current across each bond can be characterized by
\begin{equation}
\lim_{t \to \infty} \frac{\langle \mathbf{J}^2 \rangle}{t} = \frac{\sigma(\rho^*)}{L}.
\end{equation}
At the macroscopic level, this motivates adding to $\Jhat(\mathbf{x},\tau)$ a term representing Gaussian white noise with variance $\sigma [\rho(x,t)]$, which leads to an LDP for the joint probability of seeing a particular density and current profile having the form 
\begin{equation}
P[\rho(\mathbf{x},\tau),\Jhat (\mathbf{x},\tau)] \sim \exp \left\{ -L^d \int_{0}^{T} \int_0^1 \frac{(\Jhat (\mathbf{x},\tau) +D[\rho(\mathbf{x},\tau)]\mathbf{\nabla} \rho(\mathbf{x},\tau))^2}{2\sigma[\rho(\mathbf{x},\tau)]} \, d\mathbf{x} \, d\tau \right\}
\end{equation}
with $\Jhat$ and $\rho$ linked by the continuity equation~\eqref{e:cont}.

In principle, from here one can use the tools of variational calculus to look for the optimal density profile $\rho{(\mathbf{x},\tau)}$ leading to a given final density $\rho(\mathbf{x})$. This contraction leads to an implicit expression for the density rate function $\mathcal{F}[\rho(\mathbf{x})]$ defined in Sec.~\ref{ss:dens}.  Finding explicit solutions is a difficult task, since in general the optimal profile is time-dependent. However, successful treatments along these lines include the one-dimensional zero-range and Kipnis-Marchioro-Presutti models~\cite{Bertini02,Bertini05e}; in~\cite{Bertini02} it is also shown that the approach is consistent with the independent results of Derrida~\emph{et~al.}~\cite{Derrida01,Derrida02b} for the symmetric simple exclusion process.

It turns out to be easier to calculate the current large deviation function~\cite{Bertini06b}, in particular, if one assumes that the optimal profile leading to a particular current fluctuation is independent of time and that the optimal current is constant in space. In one dimension, the first assumption can be shown~\cite{Bertini05d} to be equivalent to the additivity principle of Bodineau and Derrida~\cite{Bodineau04}, which is known to break down in systems with dynamical phase transitions; the second assumption is important for higher-dimensional systems~\cite{PerezEspigares11}.  Under these conditions, the rate function for the (rescaled) current is then given by
\begin{equation}
I(\Jhat)=\min_{\rho(\mathbf{x})}\int_0^1 \frac{(\Jhat+D[\rho(\mathbf{x})]\mathbf{\nabla} \rho(\mathbf{x}))^2}{2\sigma[\rho(\mathbf{x})]} \, d\mathbf{x}. \label{e:dint}
\end{equation}
For any particular model, this integral must be minimized with $\rho(\mathbf{x})$ matched to the reservoir densities at the boundaries.  This generically yields a current distribution with non-Gaussian tails even though the fluctuations themselves are locally Gaussian.    

Various scenarios for the form of the macroscopic current large deviation function (including those indicating dynamical phase transitions) are discussed in~\cite{Bertini06b}.  The example models treated there include the one-dimensional zero-range process with $w_n$ unbounded, the special case $w_n=n$ corresponding again to non-interacting random walkers.  It has recently been pointed out that, if Eq.~\eqref{e:dint} holds, the optimal density profile is the same for all currents with the same magnitude $|\Jhat|$ leading to what has been dubbed an \emph{isometric fluctuation relation}~\cite{Hurtado11b}. Macroscopic fluctuation theory has also been generalized to treat models with dissipated energy \cite{prados2011}.

\section{Final remarks}
\label{seccon}

In this chapter we have merely skimmed the surface of the way in which large deviation theory can provide a framework for understanding existing results in the theory of nonequilibrium systems and probing for new ones.  We conclude here with some pointers to other relevant work and ideas for future research directions.

Firstly, we note that the study of fluctuation theorems and relations, as briefly touched on in Secs.~\ref{ssecneqld} and~\ref{ss:curr}, is a vast subject which percolates through many of the contributions in this volume; for overviews, see for example the chapters by Spinney and Ford, Gaspard, and Rondoni and Jepps.  In particular, we have not discussed here the important concept of entropy production and its distribution, which is central for many fluctuation relation statements and can often be expressed in a large deviation form.

A second topic worth mentioning is the extension of the concept of statistical ensemble, discussed for equilibrium systems in Sec.~\ref{sseqld}, to nonequilibrium systems. Just as one distinguishes equilibrium systems with fixed energy (microcanonical ensemble) from systems with the energy fixed on average via a conjugate Lagrange parameter (canonical ensemble), one can construct a microcanonical ensemble of trajectories for a nonequilibrium system that is constrained to realise a particular observable value (e.g., a particular current) or a canonical ensemble of trajectories that realise that constraint on average. An example of the former ensemble, obtained for the ASEP on a ring conditioned on enhanced flux, has recently been analyzed in~\cite{Popkov10d}. A study of canonical-type nonequilibrium ensembles, which are also known as \emph{biased ensembles}, can be found in the work of Sollich and Jack~\cite{Jack10}.  In the context of large deviation theory, these ensembles can be understood in terms of conditional LDPs and the so-called Gibbs conditioning principle \cite{dembo1998}.

Related to the topic of nonequilibrium ensembles is the issue of determining configurations or states giving rise to fluctuations. We can already get information about the most probable (typical) way to realise a given current fluctuation from the eigenvector corresponding to the principal eigenvalue of the tilted generator~\eqref{e:currgen}. More work is still needed to understand the properties and correlations of such current-carrying states and constrained states in general. The associated inverse problem of determining microscopic rates which are most likely to yield given macroscopic properties has been studied by Evans~\cite{evans2004,evans2005a} and Monthus~\cite{Monthus11b}.

Throughout this chapter we have assumed that non-equilibrium systems of interest are modelled by Markov processes.  However, the memoryless property may be an inappropriate approximation for the description of many systems where long-range temporal correlations are known to be important; see e.g.,~\cite{Rangarajan03} and references therein.  Recent work to characterize the large deviation properties of certain classes of history-dependent models can be found in~\cite{Me09} and~\cite{Maes09b}.  In particular, it is shown in~\cite{Me09} that modifying a continuous-time random walker (as introduced in Sec.~\ref{ss:rw}) so that the hopping rates at time $t$ depend on the average current up to time $t$ can lead to an altered ``speed'' (i.e., power of time $T$) in the LDP for current.  Fluctuation relations with the right-hand side of~\eqref{e:MGCFT} replaced by $e^{T^{\alpha} c  m}$ have also appeared in the context of anomalous dynamics; see the contribution by Klages~\emph{et al.}\ in this book.  There is much scope for future work investigating many-particle non-Markovian processes and establishing a common framework for the results.  In this regard, and in a more general way, we expect the large deviation formalism to continue playing an important role in quantifying nonequilibrium fluctuations in small systems.

\subsection*{Acknowledgments}

HT is grateful for the hospitality of the \'Ecole Normale Sup\'erieure of Lyon and the National Institute for Theoretical Physics at Stellenbosch University, where parts of this chapter were written.

\bibliography{joinref}
\bibliographystyle{wivchnum}

\end{document}